\documentclass[prl,twocolumn]{revtex4}
\usepackage{amsmath}
\usepackage{amssymb}
\usepackage{graphicx}

\begin{document}
\title[]{The Local Field Factor and Microscopic Cascading: A Self-Consistent Method Applied to Confined Systems of Molecules}
\author{Nathan J. Dawson\footnote{Corresponding
author: dawsphys@hotmail.com} and James H. Andrews}
\address{Department of Physics and Astronomy, Youngstown State University \\ Youngstown, OH 44555}

\begin{abstract} We use a simplified self-consistent method to address nonlinear-optical cascading phenomena, which shows added microscopic cascading contributions in high-ordered nonlinear susceptibilities through fifth order. These cascading terms in the microscopic regime encompass all possible scalar cascading configurations. The imposition of geometric constraints further influences the predicted cascading contributions and  opens up additional design parameters for nonlinear-optical materials. These results are used in approximating the effective fifth-order susceptibility in thin films of C$_{60}$ monomers of varying thickness and concentration. This paper contains the corrections to the original paper that appeared in the Journal of Physics B as reflected in the content of the corrigendum that followed.
\end{abstract}

\maketitle

\section{Introduction}
\label{sec:introcasc}

Materials with large nonlinear-optical responses are used for creating devices for harmonic generation, spontaneous down conversion, the optical Kerr effect, etc.\cite{ward65.02,orr71.01,boyd92.01} Cascading (at the molecular level) has been a recent topic of interest due to the apparent increase in higher-order nonlinear optical susceptibilities that stem from products of lower-order nonlinearities.\cite{andre91.01,stege93.01,stege96.01,baev10.01}

Dolgaleva, \textit{et al}., showed that local field corrections based on Bloembergen's method \cite{bloem96.01} can predict trends in the nonlinear susceptibilities as functions of concentration.\cite{dolga07.01,dolga09.01} Here, we use a self-consistent method to approximate the local field factors and the cascading contribution of the first four hyperpolarizabilities of a geometricly-constrained system of molecules. The self-consistent method has previously been shown to give the exact solution for the second hyperpolarizability (third polarizability) in a system of two interacting dipolar molecules.\cite{dawson11.02,dawson11.03} There are many advantages of manipulating the nonlinear-optical response microscopically, such as prolonging the stability of optical solitons,\cite{desya05.01} or adding new design parameters to solid film optical limiters.\cite{kumar07.01} To this end, we introduce a self-consistent theory of microscopic cascading for a system with a large number of molecules subject to simplifying constraints to focus attention on only the most relevant tensor components in the dipole approximation. This approach is well suited for a gas-lattice model, which we use to reduce the translational degrees of freedom when calculating large-scale molecular interactions.\cite{lebwo72.01,priez01.01,romano86.01}

We then describe the effects of a confined system geometry in order to highlight key design parameters for materials composed of molecules that have a large nonlinear-optical response. By increasing the incident surface-to-volume ratio of thin samples containing molecules with a large linear- and nonlinear-optical response, higher-ordered responses can be enhanced via an increase in the linear local field and microscopic cascading. With recent advances in layered polymeric systems,\cite{singer08.01,song09.01,ponti10.01,ponti10.02} multi-layered materials could be created with enhanced nonlinear-optical properties due to surface effects caused by interleaving materials with small polarizabilities placed between each active layer.

\section{Theory}
\label{sec:theory}

\subsection{Local field effects}
\label{sub:twoaligneddips}

Consider a system of dipoles that are polarized along one molecular axis in the direction of an applied electric field, which is taken to be the $z$-axis. The components of the molecular susceptibility in the direction perpendicular to the polarizable axis are assumed to be negligible. In this system, an induced dipole, $\mathbf{p}_{i}$, is polarized by the applied field, $\mathbf{E}_a$, and the induced electric field of the surrounding molecules, $\sum\mathbf{E}_j$.

The problem becomes complicated with five degrees of freedom, two rotational and three translational. For simplicity, the molecules are assumed to be located at points on a cubic lattice.\cite{dawson11.02,dawson11.03} When the only interaction is from nearest neighbors (nn), the electric fields inside the lattice from neighboring molecules are simply $E_\perp = -p/r^3$ from a neighboring molecule in the direction perpendicular to the applied field, and $E_\| = 2p/r^3$ from a neighboring molecule in the direction parallel to the applied field.

More interactions can be included in the same fashion for second nn, third nn, and so forth. The dipole field in vector form,
\begin{equation}
\mathbf{E}_{j} = \frac{3\left(\hat{r}-\hat{r}_{j}\right) \left[\mathbf{p}_{j}\cdot \left(\hat{r}-\hat{r}_{j}\right)\right] - \mathbf{p}_{j}}{\left|\mathbf{r}-\mathbf{r}_{j}\right|^3} - \frac{4\pi}{3}\mathbf{p}_{j} \delta\left(\mathbf{r}-\mathbf{r}_{j}\right),
\label{eq:labframeE}
\end{equation}
is used to describe the field from neighboring molecules that are not translated either perpendicular or parallel to the molecule with respect to the applied field.\cite{jacks96.01}

The magnitude of the lattice local field factor for two interacting molecules, $L_{\mathrm{lat},1}$, has previously been derived, and given by
\begin{equation}
L_{\mathrm{lat},1} = \left(1-f_{1,2} \frac{\alpha}{r^3}\right)^{-1} ,
\label{eq:localf2}
\end{equation}
where $f_{1,2}$ is a geometric dependent coefficient for two identical molecules, $\alpha$ is the polarizability, and $r$ is the distance of separation.\cite{dawson11.02,dawson11.03} Note that $r = \left|\mathbf{r}_2 - \mathbf{r}_1\right|$ because $p_1$ and $p_2$ are nearest neighbors. Equation \ref{eq:localf2} does not include the Lorentz local field, and was derived from the self-consistent equation of two identical interacting dipoles subject to an applied field as opposed to the macroscopic field such that
\begin{equation}
p_1 = \sum_n k^{\left(n\right)} \left(E_a + f_{1,2} \frac{p_2}{r^3}\right)^n ,
\label{eq:polarizabilitysum}
\end{equation}
where $k^{\left(1\right)} = \alpha$ and $p_1 = p_2 = p_i$.

For an infinite isotropic volume subject to a macroscopic field, $E_M$, a spherical molecule's direct electric field in a cubic lattice is in the direction of $E_M$ such that the direct field, $E_{\mathrm{dir}}$, was shown by Lorentz\cite{lorent52.01} to be
\begin{equation}
E_{\mathrm{dir}} = \frac{4\pi}{3} \frac{N p}{V} = \frac{4\pi}{3} \frac{p}{r^3},
\label{eq:elecselfaction}
\end{equation}
where $r$ is the lattice constant and $V = N r^3$ is the volume with $N$ denoting the number of induced dipoles. When aberrations in spherical symmetry are present or a simple cubic lattice structure is not an accurate physical description, then the direct field is defined as
\begin{equation}
E_{\mathrm{dir}} = f_{i,i} \frac{p}{r^3} ,
\label{eq:genselfaction}
\end{equation}
where $f_{i,i}$ is the direct field coefficient. Note that when defining the polarizability and susceptibility in terms of the applied field, we may sum the fields produced by all other molecule, which includes the contribution from the depolarization field and $E_{\mathrm{dir}}$ in the macroscopic approach.

Assuming that each dipole in a many-body system can be approximated as equal and subject to an applied field, we can generalize Equation \ref{eq:polarizabilitysum} to give
\begin{equation}
p_i = \sum_n k^{\left(n\right)} \left(E_a + f_{i}^{\left(N-1\right)} \frac{p_i}{r^3}\right)^n ,
\label{eq:polarizabilitygen}
\end{equation}
where the effective polarizability for each molecule in the cubic lattice with $N$ lattice points is calculated by summing over the interactions of all other molecules, $j\neq i$, which gives the interaction coefficient of the $i$th molecule,
\begin{equation}
f_{i}^{\left(N-1\right)} = \sum_{j\neq i} f_{i,j} .
\label{eq:interactioncoefficient}
\end{equation}

\begin{figure}[t]
\includegraphics[scale=1]{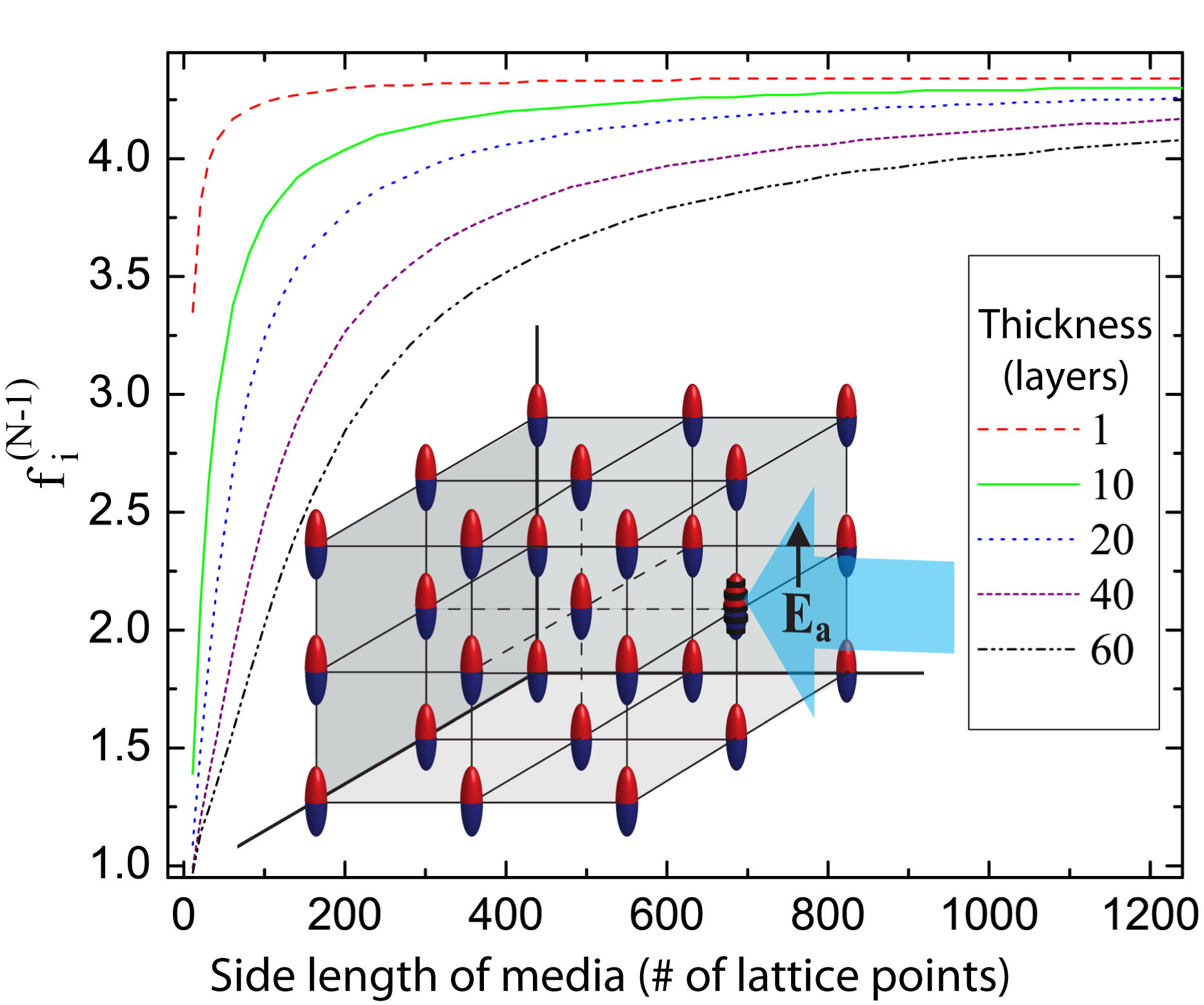}\centering
\caption{The interaction coefficient, $f_{i}^{\left(N-1\right)}$, of a molecule at the center of the surface of a film as a function of square side length for various film thicknesses. The inset shows the cubic lattice structure, where the dark striped molecule is the molecule of interest.}
\label{fig:surfgraph}
\end{figure}

\begin{figure}[t]
\includegraphics[scale=1]{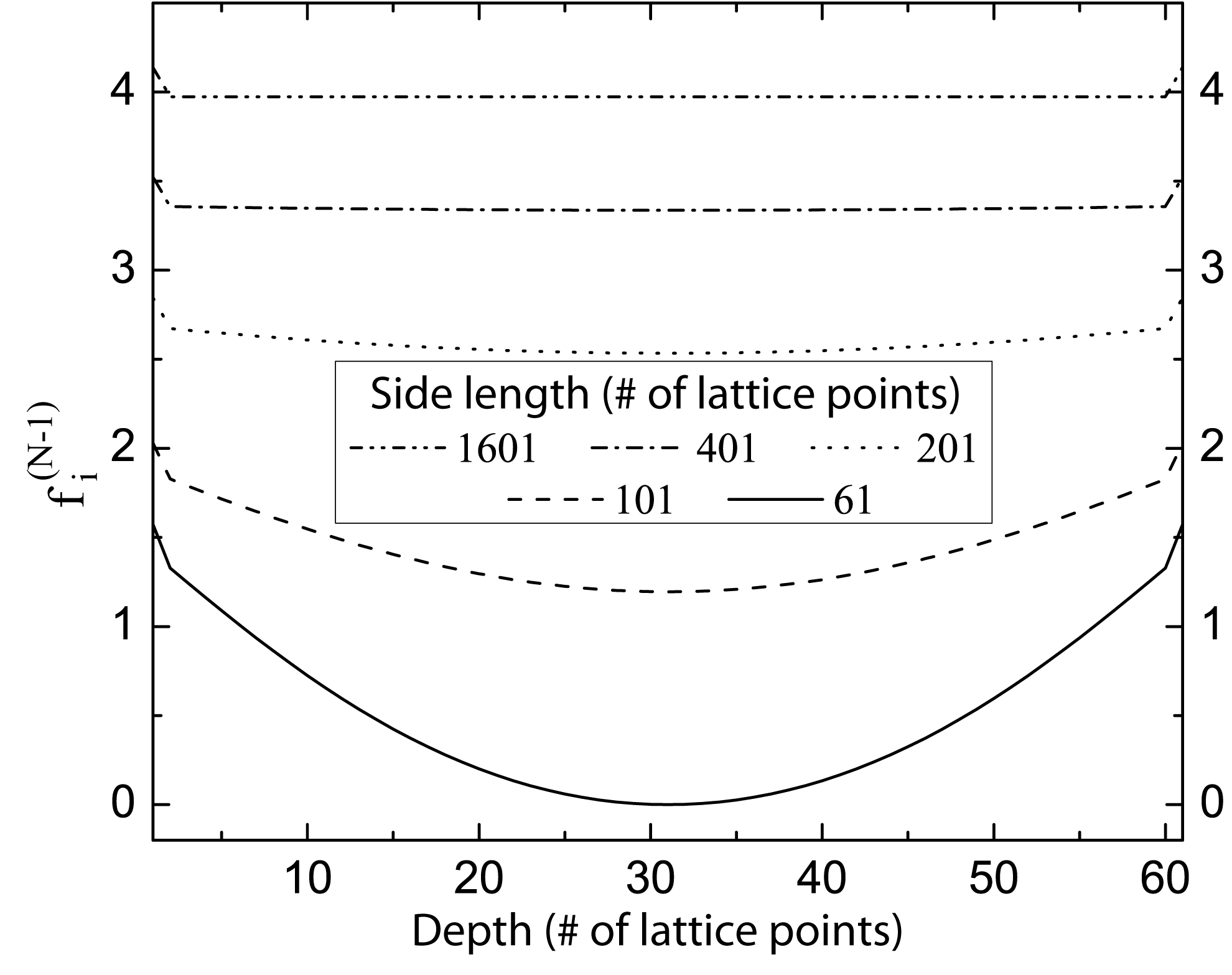}\centering
\caption{The interaction coefficient, $f_{i}^{\left(N-1\right)}$, as a function of depth in a $61$-molecule-thick film for various side lengths of square films. Note that the molecules are constrained to polarize only along an axis that is parallel to the applied field.}
\label{fig:latticegraph}
\end{figure}

Figure \ref{fig:surfgraph} shows the calculated $f_{i}^{\left(N-1\right)}$ for the gas-lattice model in the simple case where the $i$th molecule is located at the center of the surface of a square film. Though limited to this single location, the plot illustrates the rate of convergence of the interaction coefficient as a function of side length for a molecule at the surface and shows that this rate depends on the material's ratio of thickness to side length. Figure \ref{fig:latticegraph} illustrates how $f_{i}^{\left(N-1\right)}$ of a centered non-surface molecule varies as a function of depth through a $61$-molecule-thick square film for squares with various side lengths.

We define the effective (hyper)polarizabilities, $k_{\mathrm{eff}}^{\left(n\right)}$, in terms of an expansion in the applied field, $E_a$. These (hyper)polarizabilities contain terms that are less than or equal to the order in the power series for molecules with no permanent dipole moment,\cite{dawson11.02} as given by Equation \ref{eq:polarizabilitysum}. Therefore, if we assume molecules with no permanent dipole moment, $k^{\left(0\right)} \approx 0$, and choose to solve only for the linear polarizability, $\alpha_{\mathrm{eff}} = k_{\mathrm{eff}}^{\left(1\right)}$, then all higher-order terms will not be included in the solution. We can obtain a solution to the $n$th-order effective (hyper)polarizabilities by solving for the dipole moment in Equation \ref{eq:polarizabilitysum}, and then make the substitution
\begin{equation}
k_{\mathrm{eff},i}^{\left(n\right)} = \frac{1}{n!}\frac{\partial^n p_i}{\partial E_{a}^{n}} ,
\label{eq:knsolution}
\end{equation}
Thus, the effective linear polarizability of the $i$th molecule in a system of $N$ molecules that are subject to an applied electric field is
\begin{equation}
\alpha_{\mathrm{eff},i} = \alpha L_{i},
\label{eq:effalphaNmol}
\end{equation}
where the net local field factor is given by
\begin{eqnarray}
L_{i} &=& \left(1 - f_{i}^{\left(N-1\right)} \frac{\alpha}{r^3}\right)^{-1} \nonumber \\
&=& \left(1 - f_{i}^{\left(N-1\right)} \frac{N \alpha}{V}\right)^{-1} .
\label{eq:localfN}
\end{eqnarray}

The volume of a primitive cell, $V_p = r^3$, is the cube of the distance between two lattice points in our cubic gas-lattice model and summing up all cells gives $V = N V_p$, where $V$ is the volume. Thus, we can rewrite the local field factor as
\begin{equation}
L_{i} = \left(1 - f_{i}^{\left(N-1\right)} \chi^{\left(1\right)} \right)^{-1} ,
\label{eq:locfieldfacchi}
\end{equation}
where $\chi^{\left(1\right)}$ is the ``undressed'' linear susceptibility, $\chi^{\left(1\right)}=N\alpha/V$ .($\chi^{\left(1\right)} L$ is commonly known as the ``dressed'' linear susceptibility when $L_i = L_j = L$ for all $N$ molecules. The denominator in Equation \ref{eq:locfieldfacchi}, can be expanded such that $L_{i} \approx 1 + f_{i}^{\left(N-1\right)} \chi^{\left(1\right)} + \cdots$, where the expansion is typically truncated to first order since $f_{i}^{\left(N-1\right)} \chi^{\left(1\right)} \ll 1$.) Note that the local field factor in the gas-lattice model is equal to the Lorenz field factor when $f_{i}^{\left(N-1\right)} = 0$ and written in terms of the macroscopic field for a large sphere, where the depolarization field is equal and opposite of the direct field. This is also observed in the center of a cube of $61\times 61\times 61$ lattice points illustrated by the solid line in Figure \ref{fig:latticegraph}.

\subsection{Microscopic cascading}
\label{sub:microcasc}

So far our study has been limited to linear optics. Several groups have reported on the enhancement of third- and fifth-order susceptibilities from the microscopic cascading of lower-order nonlinear interactions.\cite{khurg97.01,duan03.01,dolga08.01,dolga09.02,boyd09.01} Although the molecules in this study are assumed to have no permanent dipole moment, an asymmetric potential is still possible, which allows even-ordered polarizabilities to exist. For example, the hyperpolarizability, $\beta = k^{\left( 2 \right)}$, can be nonzero without the existence of a permanent dipole moment for cases where there exists a nonzero octupole moment.\cite{ray04.01}

The nonlinear-optical response of a system is characterized by taking into account higher orders of $n$ in Equation \ref{eq:knsolution}. The (hyper)polarizabilities of the $i$th molecule, which is only allowed to be polarized in one direction with all dipole moments approximated as equal in magnitude, can be derived by solving the self-consistent dipole moment equation. After algebraic simplifications, the resulting first five effective (hyper)polarizabilities, $k_{\mathrm{eff}}^{\left(n\right)}$, are given as
\begin{eqnarray}
\alpha_{\mathrm{eff},i} &=& L_{i} \alpha , \label{eq:paraalphatwodips} \\
\beta_{\mathrm{eff},i} &=& L_{i}^{3} \beta , \label{eq:parabetatwodips} \\
\gamma_{\mathrm{eff},i} &=& L_{i}^{4} \gamma + 2 L_{i}^5 F_{i} \beta^{2} , \label{eq:paragammatwodips} \\
\delta_{\mathrm{eff},i} &=& L_{i}^{5} \delta + 5 L_{i}^{6} F_{i} \beta \gamma + 5 L_{i}^{7} F_{i}^{2} \beta^{3}  , \label{eq:paradeltatwodips}
\end{eqnarray}
and
\begin{eqnarray}
\epsilon_{\mathrm{eff},i} &=& L_{i}^{6} \epsilon + 3 L_{i}^{7} F_{i} \left(\gamma^{2} + 2\beta \delta\right) \nonumber \\
&+& 21 L_{i}^{8} F_{i}^{2} \beta^2 \gamma + 14 L_{i}^{9} F_{i}^{3} \beta^{4} ,
\label{eq:paraepsilontwodips}
\end{eqnarray}
respectively. The local field factor, $L_i$, in Equations \ref{eq:paraalphatwodips}-\ref{eq:paraepsilontwodips} is given by Equation \ref{eq:locfieldfacchi}, and $F_{i}$ is defined as
\begin{equation}
F_{i} = \frac{N}{V}f_{i}^{\left(N-1\right)} .
\label{eq:Ffactor}
\end{equation}
Equations \ref{eq:paradeltatwodips} and \ref{eq:paraepsilontwodips} are given for the first time to our knowledge, where all microscopic cascading terms represent the possible macroscopic cascading schemes.

The expressions for the polarizability and first hyperpolarizability do not contain any cross-terms. The second hyperpolarizability and higher-ordered hyperpolarizabilities, however, have cross-terms that are ordered by powers of $F_{i}$. Because $F_{i}$ is parameterized by the molecular interaction coefficient, $f_{i}^{\left(N-1\right)}$, we find that the cross-terms are a fundamental part of a finite volume due to the system with boundaries. In other words, when the finite volume has a molecular interaction coefficient of zero magnitude, the cascading contribution is zero, which gives zero contribution from a molecule's hyperpolarizability to another molecules second hyperpolarizability. The most interesting aspect of this calculation is the possibility of increasing this dependence by maximizing the sum of all interaction coefficients, $\sum f_{i}^{\left(N-1\right)}$, through geometric constraints.

\section{The enhanced nonlinear response of C$_{60}$}
\label{sec:C60}

The implications of the model in the previous section can be illustrated in a simple analysis of spherical molecules. For example, the molecule C$_{60}$ is a centrosymmetric molecule with an approximately spherical shape.\cite{hoshi95.01} In the previous sections, a planar geometry increased the molecular interaction coefficient, thereby strengthening the local field factors and cascading contributions. Therefore, a thin sheet of C$_{60}$ fullerene is proposed. C$_{60}$ is ideal for our theoretical treatment because optical poling is not required to get the maximum response due to the molecular configuration,\cite{dawson11.03,kuzyk89.03} and all cross-terms with an even-ordered molecular susceptibility, such as $\beta$ or $\delta$, are zero from centrosymmetry.

A thin film of a molecule with a large fifth-order response, such as C$_{60}$, would greatly increase a molecule's $L_i$ and $F_i$. The nonlinear properties of C$_{60}$ have been well documented including enhancements to the fifth-order nonlinear susceptibility.\cite{koudo96.01,kafaf91.01,kafaf92.01,zhang95.01,ganee09.01} The dependence on the third susceptibility squared was shown by Dolgaleva, \textit{et al}.\cite{dolga07.01,dolga09.01} Concentration dependence due to aggregation was shown to level-off the third susceptibility.\cite{blau91.01} Knize used C$_{60}$ to show how the model of free electrons confined to a sphere fit experimental results.\cite{knize04.01} Moreover, it was recently shown that large nonlinear responses were obtained from thin films of C$_{60}$.\cite{ryasn05.01} Because ``undressed'' $\chi^{\left(5\right)}$ effects are much smaller than lower-order material responses, one would conclude that the large nonlinear response of C$_{60}$ in a thin film would be an ideal test case to maximize both the local field and microscopic cascading contributions.

\begin{table}[t]\centering
\caption{Calculated Molecular Susceptibilities of C$_{60}$} 
\centering
\begin{tabular}{c c c c l} 
\hline\hline 
Symbol & \hspace{1.5cm} & Value & \hspace{1.5cm} & Units \\ [0.5ex] 
\hline
$\alpha$ &  & $1.85\times 10^{-23}$ &  & cm$^{3}$\\
$\gamma$ &  & $3.41\times 10^{-35}$ &  & erg$^{-1}$cm$^{5}$\\
$\epsilon$ &  & $6.32\times 10^{-45}$ &  & erg$^{-2}$cm$^{7}$\\
\hline
\end{tabular}
\label{table:C60values}
\end{table}

The molecular susceptibilities shown in Table \ref{table:C60values} are found by approximating C$_{60}$ as a three level model,\cite{kuzyk05.02,perez08.01} where C$_{60}$ has two large oscillator strengths with respect to their corresponding transition energies, which are much larger than the oscillator strengths associated with other transitions. We stress that the values are estimated from a 3-level model where higher order hyperpolarizabilities may be overestimated. The values of the transition energies and oscillator strengths were taken from the experimental values given by Leach, \textit{et al}.\cite{leach92.01,bulga92.01}

\subsection{Interaction coefficient components and fringe effects}
\label{sub:fringeeff}

Until now, the self-consistent model derived in the Theory Section has only been used to calculate the effective molecular susceptibilities of molecules that are constrained to polarize along a single direction. C$_{60}$, however, is a spherical molecule that will equally polarize in three dimensions. With this, we define the interaction coefficient in three dimensions, where $f_{i,x}^{\left(N-1\right)}$, $f_{i,y}^{\left(N-1\right)}$, and $f_{i,z}^{\left(N-1\right)}$ are the respective interaction coefficients in the $x$, $y$, and $z$ directions. Although the interaction coefficients are expected to be small in the directions perpendicular to the electric field, they may not be negligible in certain geometric configurations of molecules with large values of $\alpha/r^3$.

\begin{figure}[t]
\includegraphics[scale=1]{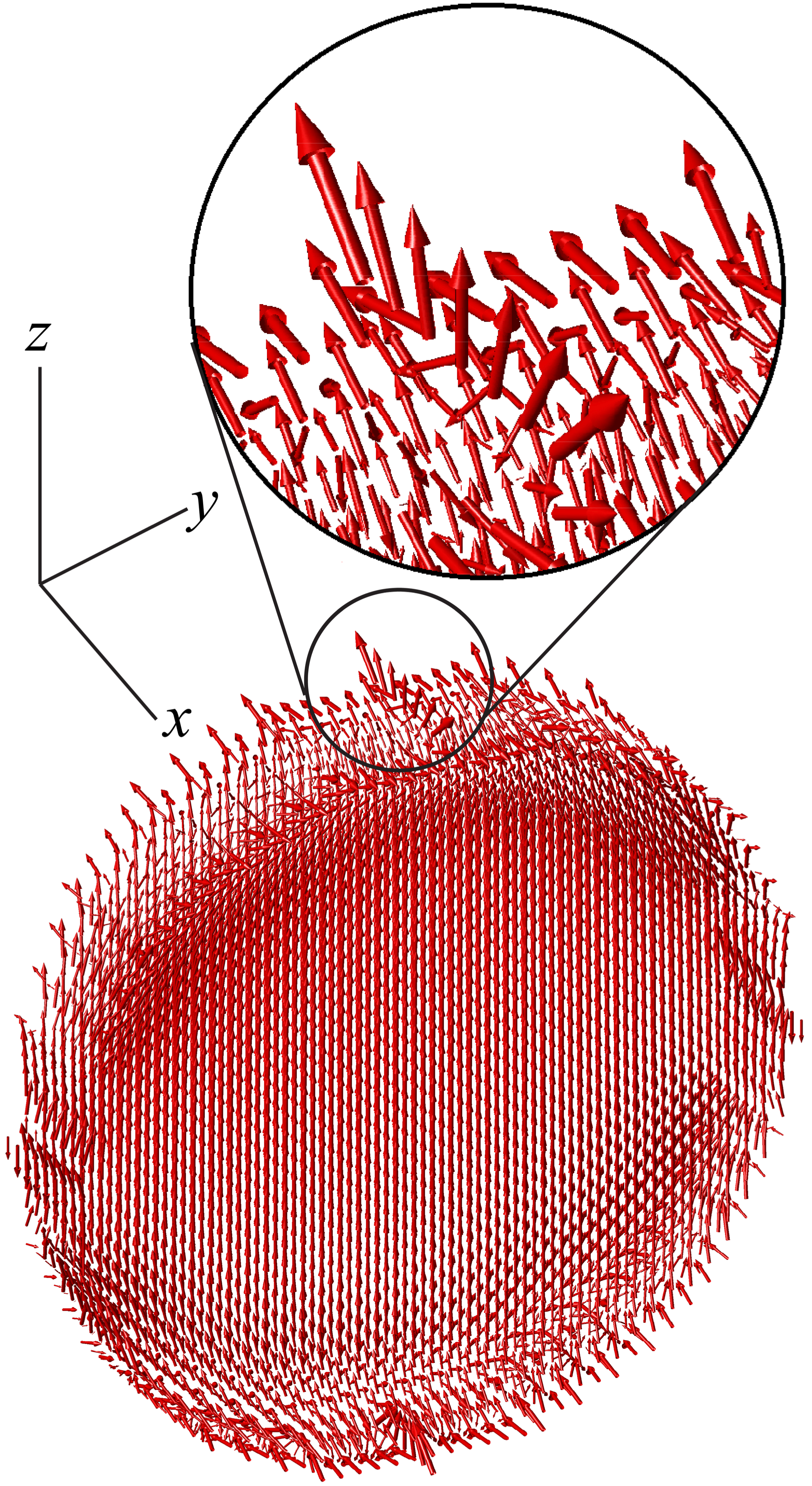}\centering
\caption{A vector diagram of the three component interaction coefficient. The interaction coefficients are the results of an off-resonant top hat beam with a cross section of 41 molecules along the $x$- and $y$-axes, where the beam is illuminating a 7-molecule thick film.}
\label{fig:vecplot}
\end{figure}

Previously, we denoted $f_{i}^{\left(N-1\right)}$ as the interaction coefficient. However, this is the interaction coefficient when all induced dipole moments and all interaction coefficients are equal in both direction and magnitude. Because the interaction coefficient changes as a function of location on the lattice, the induced dipole will also change with respect to a molecule's position on the lattice. We will denote these interactions as second-order interactions (the induced dipole from all other dipole fields that occur from first-order interactions). An example is the case where the dipole moment of one molecule is subject to the dipole fields of the other molecules but not subject to the applied field, and the second-order interaction is when the dipole fields from the molecules outside the applied field contribute to the induced dipole moment of the molecules inside the applied field.

Because the largest field in a lattice of C$_{60}$ molecules is produced by the linear polarizability with a magnitude given in Table \ref{table:C60values}, and the distance of closest approach, $r_{\mathrm{min}}$, is approximately $7.1\,$\AA,\cite{casti94.01} we find that $\alpha/r_{\mathrm{min}}^3 \approx 0.05$. Therefore, all second-order interactions will have a contribution that scales as $\alpha^2/r^6 \approx 2.5\times10^{-4}$. Thus we choose to neglect second-order interactions in the C$_{60}$ lattice based on the small $\alpha^2/r^6$ dependence. To this end, we focus on the nonlinear susceptibility from only the applied field and first-order interactions, and assume that $p_i \approx p_j$ through the region illuminated by a beam with constant magnitude in that region. Inasmuch as the applied field is to be taken along the $z$-axis, $p_{i,z}$ is the only non-zero component of the induced dipole moment produced by the applied field. Due to the possibility of a anisotropic system geometry, however,  all three components of $\mathbf{p}_i$ may exist due to orientations of a dipole field from a molecule in the system with respect to other molecules. Although small enough for a valid first-order interaction approximation, components of the interaction coefficient in the directions perpendicular to the applied field are larger near the surfaces of illuminated regions. This fringe behavior is illustrated in the blowup of the vector diagram of the interaction coefficients shown in Figure \ref{fig:vecplot}.

\subsection{Thin films subject to an off-resonant top hat beam}
\label{sub:fifthenhance}

The effective fourth hyperpolarizability, $\epsilon_{\mathrm{eff}}$, of the centrosymmetric molecule C$_{60}$ depends on the polarizability, second hyperpolarizability, fourth hyperpolarizability, C$_{60}$ concentration, and geometric constraints. The polarizability, second hyperpolarizability, and fourth hyperpolarizability are assumed to be constants of C$_{60}$. This is true so long as we ignore perturbations of the energy levels due to molecular interactions such as aggregation effects. With this assumption, the two experimentally controllable parameters are the concentration and geometry. For this statistical model, we wish to look at the fifth-order susceptibility of a thin film of C$_{60}$ and vary the concentration and thickness. For the sake of comparative analysis, the coherent beam's diameter is set to be a function of concentration so that there is always a constant area of molecules being illuminated.

The effective fifth-order susceptibility along the direction of the electric field for molecules that possess a nonzero value for both the second and fourth hyperpolarizabilities is
\begin{eqnarray}
\chi_{\mathrm{eff}, zzzzzz}^{\left(5\right)} &=& \frac{\epsilon}{V} \sum_{i}^{N} L_{i}^{6}  + 3 N \frac{\gamma^2}{V^2} \sum_{i}^{N} L_{i}^{7} g_{i}
\label{eq:chi5}
\end{eqnarray}
where $g_i = f_{i}^{\left(N-1\right)}$. The summations in Equation \ref{eq:chi5} are functions of the concentration and geometry. Therefore, when ignoring energy perturbations, the effective fifth-order susceptibility can be maximized by adjusting these two parameters. Although small, it should also be noted that the fringe effects discussed in the previous subsection become apparent in this geometry, and will give rise to nonzero components of $\chi_{\mathrm{eff}, ijklmn}^{\left(5\right)}$ that would otherwise be zero.

\begin{figure}[b]
\includegraphics[scale=1]{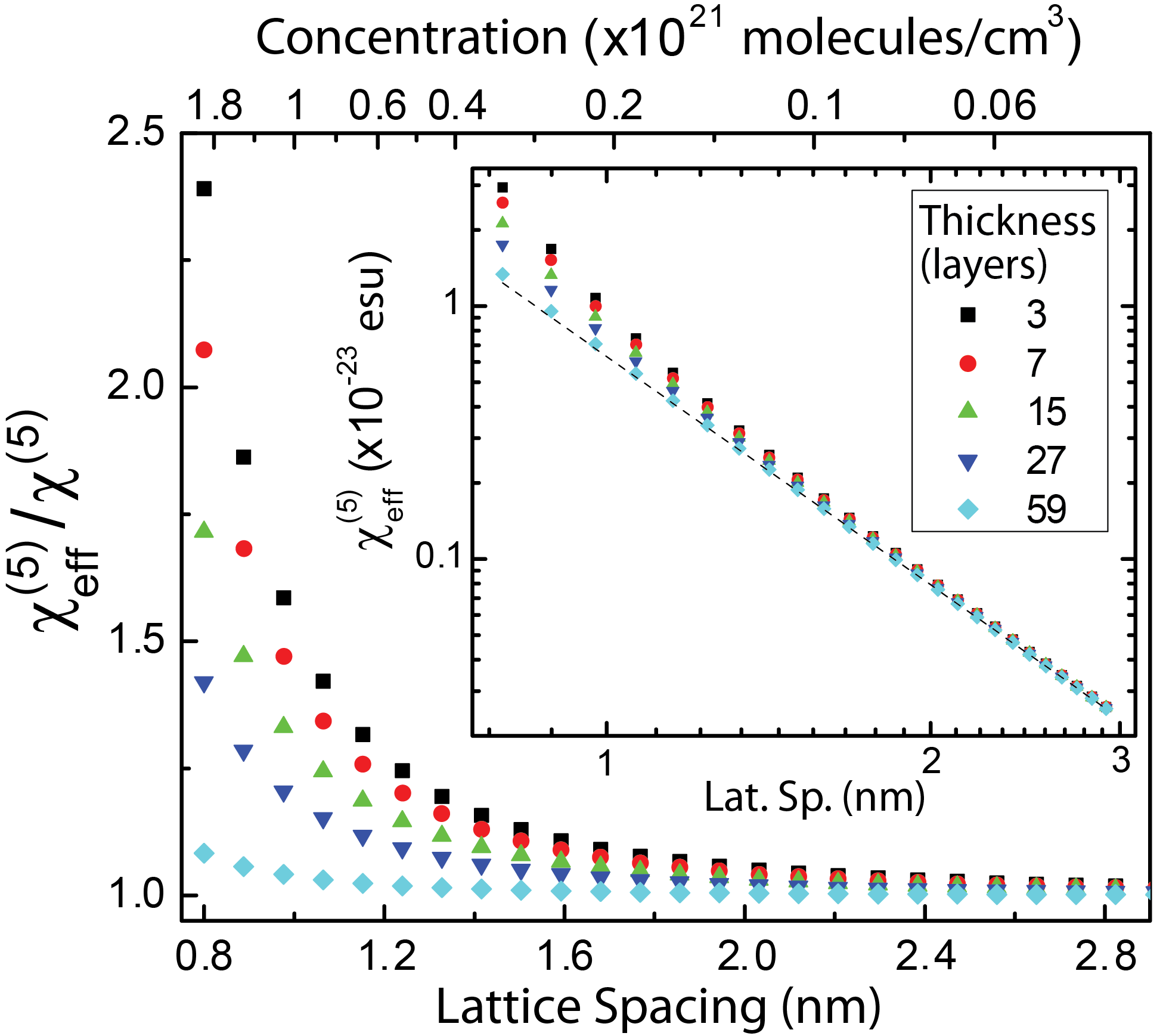}\centering
\caption{The fraction of the effective fifth-order susceptibility over the undressed fifth-order susceptibility as a function of the separation distance between molecular centers illuminated by a top hat beam with an $81\times81$ molecule cross-section. The inset shows the calculated $\chi_{\mathrm{eff}}^{\left( 5 \right)}$ as a function of the separation distance on a log-log scale. The dashed line in the inset illustrates the undressed value, which is a linear function of concentration.}
\label{fig:c60}
\end{figure}

Figure \ref{fig:c60} shows the calculated $\chi_{\mathrm{eff}}^{\left(5\right)}$ of a thin film composed of C$_{60}$ molecules subject to a normally-incident top hat beam with a frequency far below resonance, i.e., $dE_a /dx \approx 0$. Here, $\chi_{\mathrm{eff}}^{\left(5\right)}$ is calculated as a function of concentration for different film thicknesses.

The effective fifth-order susceptibility shown in Figure \ref{fig:c60} is calculated by summing the effective fourth hyperpolarizabilies of every molecule in the top hat beam radius and dividing by the cylindrical volume. Although the number of molecules in the calculation is small relative to large collimated beams, one can predict the values of $\chi_{\mathrm{eff}}^{\left(5\right)}$ for thin samples in the macroscopic regime by scaling the fraction of thickness to beam diameter, thereby strengthening the approximation of a negligible fringe effect near the beam edge.

We stress that most of the terms in Equation \ref{eq:paraepsilontwodips} do not contribute to the fifth-order susceptibility of C$_{60}$ due to centrosymmetry. One would be correct in concluding that an asymmetric molecule would greatly increase the number cascading terms, although this does not necessarily mean that the total fifth-order nonlinear response will always be larger.

\subsection{Approximations and Assumptions}
\label{sub:appandass}

The results of these calculations are based on a number of approximations. The model does not account for aggregation effects since each molecule is required to be positioned on a lattice point at a fixed separation. Therefore, large energy perturbations and the potential for many molecular systems to ``level-off'' the nonlinear response at high concentrations are not considered. A complete theory of cascading contributions and the influence of molecular interactions on the transition energies has been previously derived for the two molecule system,\cite{dawson11.03} which can be folded into the current model with some additional rigor. A full tensorial treatment up to third order of the cascaded nonlinear effects with molecular correlations is in ref \cite{andre91.01}.

An off-resonant response is assumed such that the frequency of the electric field approaches zero. Thus, retardation effects over large distances as well as transitional behaviors near resonance are not treated. Also, because there are field gradients, albeit small, there will be effects from the dipole-quadrupole polarizability, dipole-octupole polarizability, quadrupole-quadrupole polarizability, etc., leading to deviations from the dipole approximation. Finally, all dipole moments are taken as equal to allow a self-consistent approach. This assumption ignores higher order interactions, which may play an important role for asymmetric systems that contain very small numbers of molecules.

As an additional point, the electric field is assumed to be uniform throughout a cylindrical volume, but varying the magnitude of the applied field over the substrate, such as a thin film subject to various beam profiles, would allow one to approximate the cascading contribution of an optical material for a multitude of experimental conditions. Under these circumstances, the dipole moments can no longer be assumed to be equal and other assumptions must be made about the spatial distribution of dipole moments induced by the applied field. For example, a Gaussian beam profile can be approximated by a Gaussian distribution of dipole moment magnitudes.

\section{Conclusion}
\label{sec:conclusion}

The self-consistent method used to evaluate the effective nonlinear susceptibilities provides a solution that reveals multiplicative factors that are the same as the Bloembergen's approach provided that the power series representation of nonlinear optics is consistent. These multiplicative factors show that the cascading contribution is not just the sum of lower order terms multiplied by a local field factor, but that some cascading contributions are larger than others. Furthermore, the microscopic cascading terms that arise from the self-consistent method encompass all possible macroscopic cascading schemes, and are ordered in terms of $F_i$.

We have shown that it is possible to increase the local field factors and cascading contributions in higher-ordered nonlinear responses by forced geometric constraints. Inasmuch as the local field factor can increase the bulk nonlinear response, a decrease in the distance of a molecule from a surface that is perpendicular to the direction of the applied field can greatly increase a molecule's interaction coefficient with respect to the surrounding molecules.

The molecule C$_{60}$ fullerene was used as a case study, which showed an increase in the effective fifth-order susceptibility via molecular interactions. Moreover, this enhancement from molecular interactions was shown to increase with a decrease in molecular separation distances. Thus, increasing the concentration will not only increase the nonlinear response of the polarization density based on number density, but also will show an increase in nonlinearity due to interactions at relatively short distances. These interactions are shown to further increase due to geometric constraints imposed on the active volume. The model explicitly shows how a material can achieve a larger effective nonlinear response than predicted for a bulk material.

\section{Acknowledgements}
\label{sec:acknowledgements}

The authors are grateful to the National Science Foundation for financial support from the Science and Technology Center for Layered Polymeric Systems under grant number DMR 0423914. The authors are also grateful for the support of the State of Ohio, Department of Development, State of Ohio, Chancellor of the Board of Regents and Third Frontier Commission, which provided funding in support of the Research Cluster on Surfaces in Advanced Materials. The authors also thank Prof. Michael Crescimanno for helpful discussions.

\end{document}